\documentclass[12pt]{article}
\usepackage{fleqn}
\usepackage{latexsym}
\usepackage[dvips]{graphicx}
\newcommand{\be}{\begin{equation}}
\newcommand{\ee}{\end{equation}}

\begin{document}

\title{\vskip-1.7cm \bf ENTRAINMENT OF AN INERTIAL REFERENCE FRAME
BY AN ACCELERATED GRAVITATING SHELL}
\author{L.I.Men'shikov$^{1}$, I.A.Perevalova$^{2,3}$ and A.N.Pinzul$^{2,3}$}
\date{}
\maketitle
\hspace{-8mm}$^{1}${\em RNC Kurchatov Institute}\\ 
\hspace*{-2mm}$^{2}${\em University of Alabama}\\
\hspace*{-2mm}$^{3}${\em Moscow Institute of Physics and Technology}
\begin{abstract}
For the example of an accelerated shell we show that omission of the
energy-momentum tensor (EMT) of the body that causes the acceleration
and the tensions due to this acceleration can lead to a paradoxical
result; Namely, the entrainment of an inertial 
frame by the accelerated shell in the direction opposite to that of the acceleration. We consider several models and demonstrate that the correct result can be obtained only if {\it all} components of the full EMT are adequately taken into account, 
and the problem statement is {\it physically} correct.
\end{abstract}

%\begin{center}
\section{Introduction}
%\end{center}

Before the theory of general relativity was conceived, Einstein published the paper {\it Is there a gravitational interaction similar to electromagnetic induction?} \cite{1}, in which the effect of an accelerated shell on a test body placed 
inside it was discussed. The problem was analyzed using semi-classical
concepts (here we use the term {\it classical} in the sense of
Newton's gravitation), and Einstein had come to the following result
(in linear approximation with respect to 
the acceleration): the test body is driven by the shell with acceleration $a_{body}=\frac{3M}{2R}a_{shell}$, where $M$ is the shell mass and $R$ is its radius.

After the formulation of the theory of general relativity, Lense and
Tirring \cite{2} considered a similar problem. They studied the entrainment of inertial reference frames by material currents on the basis of Einstein's linearized equations. The 
results were applied to analyzing a rotating shell, and they obtained the well-known solution in which the reference frame inside the shell rotates with the frequency $\omega_{i.f.}=\frac{4}{3}\frac{M}{R}\omega_{shell}$. Later Brill and Cohen 
\cite{3} calculated a metric for a rotating shell assuming a linear approximation in $\omega $, and this metric was identical to Kerr's one outside the shell. In the case $\frac{M}{R}\ll 1$, it reduced to that calculated by Lense and Thirring.

Here we analyze an accelerated shell having at our disposal Einstein's
equations. In that sense we have much more advantage than Einstein
had. Using the linear approximation of general relativity we shall consider several 
different statements of the problem. In one case, we shall demonstrate the well-known and methodologically important result that in the case of the general relativity one should take into account the full energy-momentum tensor of the all objects 
relevant to given problem. In the other case, the incorrect and mainly paradoxical results will be received. 

The paper is organized as follows. Section 2 takes into account only the EMT of the shell. Section 3 analyzes a closed system of {\it a shell + a massless string} driving the shell. It is shown that tension in the string and shell lead to the 
additional terms in the expression for the acceleration of the probe body. But in this case again we have the incorrect result. The reason for that is quite different from the first example and we shall investigate it for the methodological 
purposes. Finally, Section 4 analyses a consistent model, namely a charged shell in a DC electric field. It is shown that the EMT of the electromagnetic field also contributes to the acceleration of the reference frame.

%\begin{center}
\section{Uniformly Accelerated Shell}
%\end{center}

Let's choose a frame system $x^i=(t,x,y,z)$ build on the base ofbuilt
on "the rigid rod lattice", which is stationary with respect to the far stars. Here $t$ is time synchronized with the remote clocks. Let a rigid infinitesimally thin shell of the radius $R$ be 
accelerated along the  $z$-axis at the constant acceleration $a$. (All
limitations on $R$ and $a$ will be written below.) Now our frame is
influenced by the shell and as a result it is not inertial. A free falling probe will 
move with some acceleration ${\bf a}_{body}$ with respect to "the rigid rod lattice". We can use the system of such probe to construct an inertial frame system (obviously, it is a question of the {\it local} inertial frame systems). So we 
can say that the inertial frame systems  are intrained by the accelerated shell.

It's obvious that in the given case it is practically impossible to find the exact solution of Einstein's equations. So we shell work in the weak field limit and with all velocities less than light: $R_g \ll R$ and $ Ra \ll 1$, where 
$R_g=2M$ is the shell gravitational radius and $a$ is its acceleration ($c=G=1$). In this case, Einstein's equations are reduced to the following form \cite{4}
\be
\label{1}
\Box \bar h_{ik}=16\pi T_{ik}\ \ \ ,
\ee
where $h_{ik}=\bar h_{ik}-\frac{1}{2}\eta_{ik}\bar h$, $\bar h=\bar h_{ik}\eta^{ik}$, $g_{ik}=\eta_{ik}+h_{ik}$, and $\eta_{ik}=\mbox{diag}(1,-1,-1,-1)$.

The general solution of Eq.(\ref{1}) (which satisfies given the boundary conditions) is expressed as \cite{4}
\be
\label{2}
\bar h_{ik}({\bf r},t)=-4\int\frac{T_{ik}({\bf r'},t')}{|{\bf r-r'}|}d^3
{\bf  r'} \ \ \ ,
\ee
where $t'=t-|{\bf r -r'}|$ is "retarded time".

The points of the rigid rod frame system move along the trajectories $x^{\alpha}=const$, where $\alpha =1,2,3$. As we explained above, that motion is noninerial (nongeodesic) and characterized by the acceleration 4-vector $w^i =\frac{{\bf 
D}u^i}{dS}\approx (0,-{\bf a}_{body})$, where $u^i=\frac{dx^i}{dS}=(u^0,0,0,0)$ is the velocity 4-vector. So we have
\begin{displaymath}
({\bf a}_{body})_{\alpha}= - \frac{{\bf D}u^i}{dS}= - \Gamma^{\alpha}_{ik}u^iu^k= - \Gamma^{\alpha}_{00}(u^0)^2 \approx \Gamma^{\alpha}_{00}= \dot h_{0\alpha}- \frac{1}{2}\nabla_{\alpha}h_{00}\ \ \ .
\end{displaymath}
Here $\nabla_{\alpha}\equiv \frac{\partial}{\partial x^{\alpha}},\ \dot h_{0\alpha}\equiv  \frac{\partial h_{0\alpha}}{\partial t}$. From Eq.(\ref{2}) and the relations $h_{0\alpha} \bar h_{0\alpha},\ h_{00}=\frac{1}{2}\sum_i \bar h_{ii}$ we have
\be
\label{3}
({\bf a}_{body})_{\alpha}=-4\int\frac{\dot T_{0\alpha}({\bf r'},t')}{|{\bf r-r'}|}d^3{\bf  r'}+ \nabla_{\alpha}\Phi \ \ \ ,
\ee
where $\Phi = \int \sum_i \frac{T_{ii}({\bf r'},t')}{|{\bf r-r'}|}d^3{\bf  r'}$. It is obvious from the axial symmetry that $({\bf a}_{body})_{x}=({\bf a}_{body})_{y}=0$ and $({\bf a}_{body})_{z}\equiv a_{body}\ne 0$. Note that the expression for 
$\Phi$ includes the non-covariant summation over $i$ but not the covariant trace.

In (\ref{3}) we have the $T_{00}, T_{\alpha\alpha}$ and $T_{0\alpha}$
components of the EMT that describe mass distribution, tensions and
energy-momentum currents respectively in the all bodies. One cannot drop any of these terms without detailed consideration. 
Let's try to calculate ${\bf a}_{body}$ from the assumption that the
acceleration is determined only by the mass distribution, i.e. $T_{00}$ corresponding to the shell (which we regard as the most massive body in our system). This assumption is 
based on the Newtonian gravity and finiteness of the gravity
propagation speed only. So, let's keep only $T_{00}$ of
the shell in (\ref{3}).

In our calculations, we retain only the terms linear in $a$. This means that we should ignore relativistic effects proportional to $\frac{v^2}{c^2}\sim a^2$. In the approximation of isotropic matter distribution  the energy-momentum tensor is given 
by
\be
\label{4}
{\bf T} = (\rho + p){\bf u}\otimes {\bf u} - p{\bf g}\ \ \ ,
\ee
where $\rho$ is the density of matter and $p\sim a$ is pressure. Therefore we obtain 
\begin{displaymath}
T^{00}= (\rho + p)(u^0)^2 - pg^{00}\approx \rho (u^0)^2 \ \ \ ,
\end{displaymath}
\begin{displaymath}
u^0 = \frac{dt}{dS}\approx \frac{1}{1+\beta a_{body}z}\approx 1- \beta a_{body}z \ \ \ ,
\end{displaymath}
where $\beta$ is of order one. In the last expression we have to drop
the $\beta a_{body}z$ term because it leads to a correction of order
$\left(\frac{R_g}{R}\right)^2 a$. This is beyond the accuracy of our approximation.

Taking in account the shell acceleration and that at $t=0$ its center of mass is at rest at the origin, we have an expression for $T^{00}({\bf r},t)$
\be
\label{5}
T^{00}({\bf r},t) = \frac{M}{4\pi R^2}\delta\left[ |{\bf r}-\frac{1}{2}{\bf a}t^2| - R \right]\equiv \rho\left( {\bf r}-\frac{1}{2}{\bf a}t^2\right)\approx \rho ({\bf r})-\frac{1}{2}{\bf a}t^2\nabla\rho ({\bf r})\ \ \ .
\ee

It is obvious that the first summand does not contribute to the
acceleration $a_{body}$. (This follows from the spherical symmetry and
it can also be seen from (\ref{3}) where the differentiation with respect to $z$ would yield zero). 
By combining Eqs. (\ref{3}) and (\ref{5}), we obtain
\begin{displaymath}
a_{body} = - \frac{1}{2}{\bf a}\partial_z \int |{\bf r}-{\bf r'}|\nabla'\rho({\bf r'})d^3{\bf r'}\ \ \ .
\end{displaymath}

We are interesting in the acceleration of the free body at $t=0$ when shell comes to rest. So, everywhere and particularly in Eq.(\ref{5}), we will set $t=0$, i.e. $t'=-|{\bf r}-{\bf r'}|$. Close to the point of origin we have
\begin{displaymath}
|{\bf r}-{\bf r'}|\approx r'-\hat {\bf r} '{\bf r}\ \ \ ,
\end{displaymath}
where $\hat{\bf r} ' = \frac{{\bf r'}}{r'}$. For the same reason that we have dropped the first summand in Eq. (\ref{5}) we should omit the first term independent of {\bf r} in the last expression. Finally, we have
\begin{eqnarray}
a_{body} & = & \frac{1}{2}{\bf a}\int (\hat{\bf r} ')_z \nabla'\rho ({\bf r'})d^3{\bf r'} = - \frac{1}{2}{\bf a}\int d^3{\bf r'} \rho ({\bf r'}) \nabla' (\hat{\bf r} ')_z 
\nonumber \\
    & = & - \frac{1}{2}{\bf a}\int d^3{\bf r'} \rho ({\bf r'}) 
\left(
\frac{\hat{\bf z} '}{r'} -\frac{\hat{\bf r} 'z'}{{r'}^2}
\right)
\nonumber\ \ \ .
\end{eqnarray}

After integration over the angular variables (which is trivial owing to the spherical symmetry of the configuration) we obtain an absurd result
\be
\label{6}
a_{body} = -\frac{1}{3}\frac{Ma}{R}\ \ \ .
\ee

On the basis of general arguments one could assume that $a_{body}$ should be of the order of $(R_g/R)a$, but there is no reason whatsoever why ${\bf a}_{body}$ should point in the direction opposite to that of ${\bf a}$.

It is obvious, that the paradox appeared because our system is not closed. Bodies causing shell acceleration and tensions and energy-momentum currents in those bodies were not taken into consideration. The intuition based on the Newton's law of 
gravity obviously let us down.

Contribution of Eq. (\ref{6}) to the probe body acceleration is due to retarding effect of the gravity signals emitted by the shell. Because of analogous reason there is electric field ${\bf E}=-\frac{2q{\bf a}}{3c^2R}$ in the center of an 
accelerated sphere with charge $q$. 

%\begin{center}
\section{A Shell Driven by a String}
%\end{center}

To correct the above situation, let us start with the following ``simple'' model: the shell accelerated by an infinitely long, massless string (Fig. 1). The word ``simple'' is in quotation marks because, as will be shown below, this model lacks a clear 
physical sense: there is no such string (at least in the frame of
classical physics).

\begin{figure}[h]
\begin{center}
\includegraphics[width=0.4\textwidth,clip]{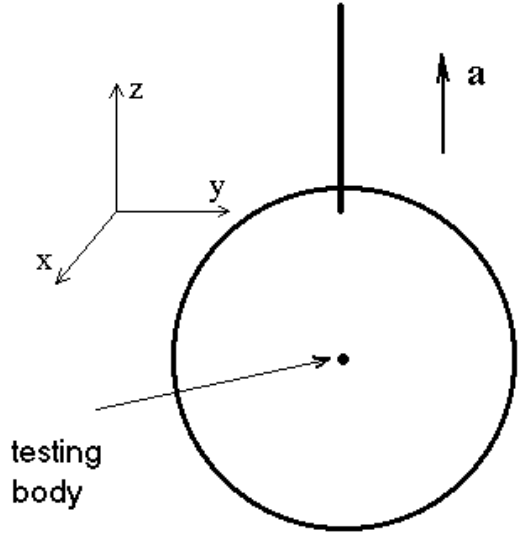}
\end{center}
\caption{Shell accelerated by an infinitely long, massless string along the $z$-axis. There is a probe, whose ``induced acceleration'' is measured at the shell center.}
\end{figure}

Now $a_{body} = a_0 + a_1 + a_2$, where $a_0$ is determined by the first summand in Eq. (\ref{3}), $a_1$ is the acceleration of the probe body calculated in the previous section considering only the shell motion, and $a_2$ takes into account all 
other factors: tensions in the shell and string, ``the spool'' on
which string is wound, and alsothe distant body that is located at a
distance $L$ from the shell and on which ``the spool'' is fixed. Let
the mass of that body along with ``the spool'' be $m$. Then its contribution to the $a_{body}$ is of the order of $m/L^2$ and one can neglect it if $L$ is large enough. So, $a_2 = a^{tension}_{2} + a^{string}_{2}$. To begin with consider the term due to tension in the shell:
\be
\label{7}
a^{tension}_{2} = \partial_z \int \frac{\sum_{\alpha}T_{\alpha\alpha}({\bf r}',0)}{|{\bf r}-{\bf r}'|} d^3{\bf r}'\ \ \ ,
\ee
where $\alpha = 1,2,3$. It was taken into account because $T_{\alpha\alpha}$ is of first order in $a$ we can put in it (and due to the same reason, in the first summand in Eq. (\ref{3}))$t'=0$.

The expression for $\sum_{\alpha}T_{\alpha\alpha}$ of shell is derived
in the Appendix:
\be
\label{8}
\sum_{\alpha}T_{\alpha\alpha}=\frac{Ma}{4\pi R}\cos \theta \delta (r-R) = aR\rho (r)\cos \theta \ \ \ .
\ee

After expanding $1/|{\bf r}-{\bf r}'|$ up to the first order in ${\bf r}$, one receives from Eqs. (\ref{7}) and (\ref{8}):
\begin{displaymath}
a^{tension}_{2} = \frac{Ma}{3R}\ \ \ .
\end{displaymath}

Comparing this result to Eq. (\ref{6}), one can see that contribution
to the acceleration of the probe  due to the shell alone is zero:
tensions in the shell have compensated the effect due to mass.

Now, let's consider $ a^{string}_{2}$. Since  the string has zero mass and is infinitesimally thin, we can write down the only nonzero component of the string EMT, taking into account the axial symmetry, as
\begin{displaymath}
T_{zz}(x,y,z) = D \theta (z-R)\delta (x) \delta (y)\ \ \ .
\end{displaymath}

The constant $D$ is calculated from Eq. (A.3) in the Appendix. Using that formula, one can receive that string acts on shell with the force
\begin{displaymath}
F_x = F_y = 0\ \ \ ,
\end{displaymath}
\begin{displaymath}
F_z = - \int T_{zz} df_z = - \int T_{zz} dxdy = - D = Ma\ \  \ .
\end{displaymath}

Thus, we have
\begin{displaymath}
T_{zz} = -aM\theta (z-R)\delta (x)\delta (y)\ \ \ .
\end{displaymath}

Now it is easy to calculate the contribution of the string EMT to $a_{body}$:
\be
\label{9}
a^{string}_{2} = -\frac{Ma}{R}\ \ \ .
\ee

The easiest way to calculate $a_0$ is to use Eq. (\ref{4}):
\begin{displaymath}
T_{0\alpha}({\bf r}, t) = (\rho + p)u_0 u_{\alpha} \approx \rho u_{\alpha} \approx -\rho ({\bf v})_{\alpha}\ \ \ ,
\end{displaymath}
\begin{displaymath}
\dot T_{0\alpha}({\bf r}, t) \approx -\rho ({\bf a})_{\alpha}\ \  \ .
\end{displaymath}
Therefore
\be
\label{10}
a_0 = 4\frac{Ma}{R}\ \ \ .
\ee

Hence the total acceleration of the probe body rested at the center of shell which is accelerated by infinitely thin massless string is
\be
\label{11}
a_{body} = \frac{3M}{R}a \equiv \frac{3R_g}{2R}a\ \ \ .
\ee

If one pays attention to the sign in Eq. (\ref{9}) which tells us that
the probe is repulsed by the stretched massless string, doubts about
Eq. (\ref{11}) appear. From Eq. (\ref{3}) and the expression for $\Phi$ one can see that this gravitational 
repulsion appears in the case
\be
\label{12}
\sum_i T_{ii} < 0\ \ \ .
\ee
For the homogeneous body this means that $\epsilon + 3p < 0$, i.e. $p
< -\frac{\epsilon}{3}$. There are no known bodies or fields with such a property.

For the usual string we have $|p| \ll \epsilon \approx \rho$, where $|p| \sim \frac{F}{d^2} = \frac{Ma}{d^2}$ and $\rho$ is string density and $d$ is its diameter. Such string accelerates a probe body with the usual Newtonian acceleration $a_N \sim 
\frac{\rho d^2}{R}$. It follows, that
\begin{displaymath} 
a_N \gg \frac{|p|d^2}{R} \sim \frac{Ma}{R}\ \ \ .
\end{displaymath}
So, in the experiment with the shell driven by a string, it would impossible to detect the effect of inertial frame entrainment because it is hidden under the much more substantial effect from the string gravitational attraction. Though one can 
speak only about the gedankenexperiment because entrainment effect is so small that it cannot be detected. For example, additional phase incursion for the laser beam past through an accelerated body would be $\sim 10^{-40}$ in typical laboratory 
experiment. 

Keeping this conclusion in mind, let us proceed to the consistent system, namely a charged shell driven by an electric field.

%\begin{center}
\section{Charged Shell}
%\end{center}

\begin{figure}[h]
\begin{center}
\includegraphics[width=0.8\textwidth,clip]{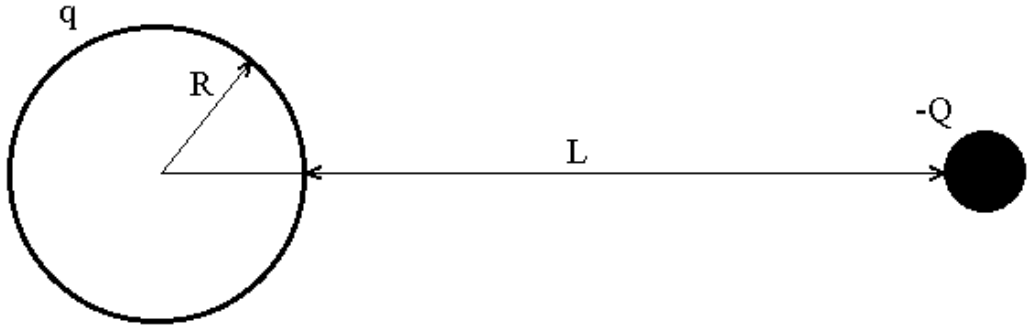}
\end{center}
\caption{A charged shell with a charge $q$ is in the field generated by the charge $-Q$. The distance between the charges is selected so that the direct effect of the mass of charge $Q$ and effects quadratic in the electric field could be 
neglected.}
\end{figure}

A shell with charge $q$ (Fig. 2) is placed in the field ${\bf E}_0$
generated by charge $-Q$. The charge is sufficiently far away (at the
distance $L$) so that one can consider the field ${\bf E}_0$ as
homogeneous and neglect the gravitational effect 
of the charge $-Q$ onto a probe. The shell acceleration is
\be
\label{13}
{\bf a} = \frac{q{\bf E}_0}{M},\ \ \ E_0 = \frac{Q}{L^2}\ \ \ .
\ee  

\begin{figure}[h]
\begin{center}
\includegraphics[width=0.4\textwidth,clip]{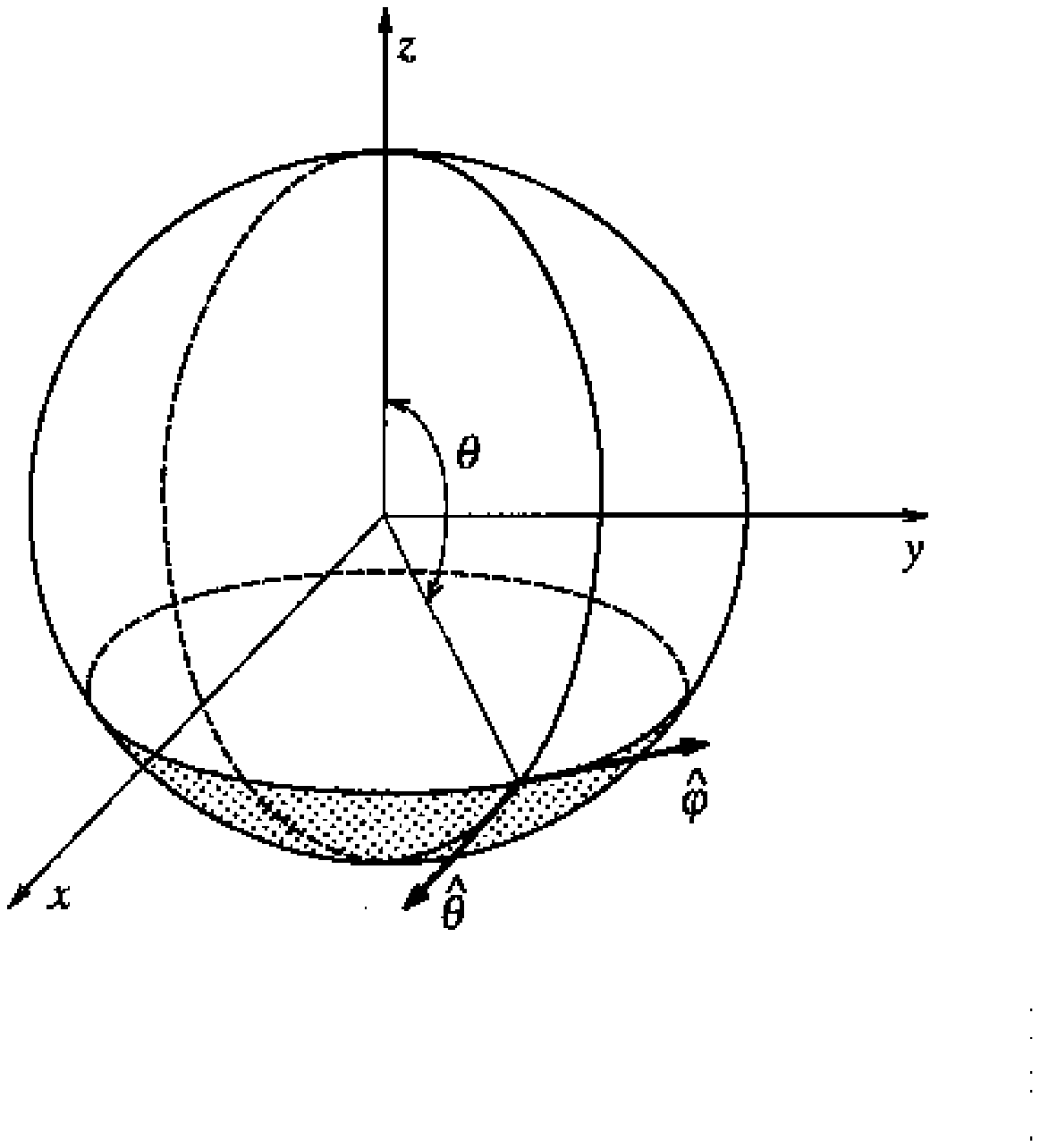}
\end{center}
\caption{Illustration to the calculation of the component $A(\theta )$ in the shell tension tensor.}
\end{figure}

Let us calculate the tension generated in the shell (Fig. 3). The charge and mass of the dashed section are  
\begin{displaymath}
q_1 = q\frac{1+\cos \theta}{2},\ \ \ M_1 = M\frac{1+\cos \theta}{2}\ \
\ ,
\end{displaymath}
respectively, where $\theta$ is the angle of spherical frame system
with a polar axis $z$ directed along ${\bf a}$. This section is acted
upon by the inertial force ${\bf F}_{in}=-M_1{\bf a}$ and
electrostatic force ${\bf F}_{el}=q_1{\bf E}_0$. Due to the axial symmetry, the tension tensor can be expressed as (see Appendix for detailed calculations)
\begin{displaymath}
T_{\alpha\beta} = \left[ A(\theta){\hat \theta}_{\alpha}{\hat \theta}_{\beta} + B(\theta){\hat \varphi}_{\alpha}{\hat \varphi}_{\beta} \right] \delta (r-R)\ \ \ ,
\end{displaymath}
where $\hat \theta ,\ \hat \varphi$ are unit vectors tangent to the sphere and directed along meridians and parallels respectively.

Then, from Eq. (A.3) and the formula $d{\bf f} = - \hat \theta r \sin \theta d\varphi dr$ follows the condition for the force balance along z-axis:
\begin{displaymath}
\frac{q{\bf E}_0(1+\cos\theta)}{2} - \frac{m{\bf a}(1+\cos\theta)}{2} + {\bf F} = 0\ \ \ ,
\end{displaymath}
where ${\bf F} = -2\pi A(\theta)R\sin^2 \theta \hat z$. It follows from this equation and Eq. (\ref{13}) that $A(\theta) = 0$.

\begin{figure}[h]
\begin{center}
\includegraphics[width=0.4\textwidth,clip]{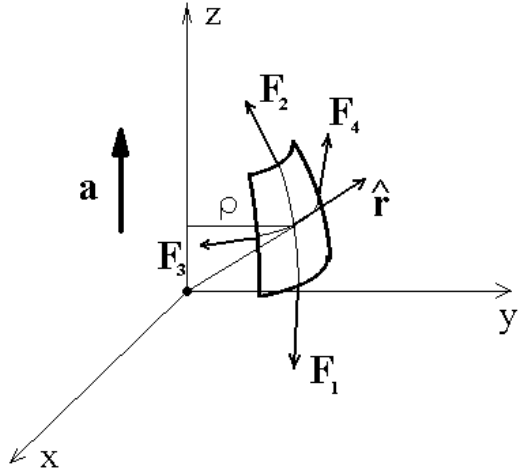}
\end{center}
\caption{Illustration to the calculation of the component $B(\theta )$ in the shell tension tensor. The forces  ${\bf F}_1$ and ${\bf F}_2$ are tangent to the meridian $\varphi = const$, and the forces ${\bf F}_3$ and ${\bf F}_4$ to the 
parallel $\theta = const$. $\rho$ is the distance between the shell element and $z$-axis.}
\end{figure}

To find $B(\theta)$, let us consider the balance condition for $r$-components (see Fig. 4).Using a relation $\Delta q{\bf E}_0 = \Delta M{\bf a}$, where $\Delta q$ and $\Delta M$ are the charge and mass of a shell element and calculations similar 
to those from Appendix, one can show that $B(\theta) = 0$. So, the
external electrical field doesn't cause tensions in the freely moving
shell. In fact, this is obvious because the electric forces are exactly balanced by the inertia forces in 
the accelerated frame system.

The contribution to $a_{body}$ due to the first summand in Eq. (\ref{3}) is again determined by Eq. (\ref{10}): presence of the field doesn't change this result. Indeed, using a conservation law for the full energy-momentum tensor $T_{ik}$, for $k=0$ 
we have 
\begin{displaymath}
{\dot T}_{0\alpha} = \frac{\partial T^{0}_{\alpha}}{\partial t} = -\frac{\partial T^{\beta}_{\alpha}}{\partial x^{\beta}} = \frac{\partial T_{\beta\alpha}}{\partial x^{\beta}}\ \ \ ,
\end{displaymath}
\begin{displaymath}
\int\frac{{\dot T}_{0\alpha}({\bf r}', t)}{|{\bf r}-{\bf r}'|}d^3{\bf r}' = \frac{\partial}{\partial x^{\beta}}\int\frac{T_{\beta\alpha}({\bf r}', t)}{|{\bf r}-{\bf r}'|}d^3{\bf r}'
\end{displaymath}
(as we said above, one should neglect signal delay).

Now, one should set $t=0$ in Eq. (\ref{3}), leave only the field
contribution and neglect terms of the order of $v^2/c^2$, particularly
magnetic field. The electrostatic field inside stopped shell is zero, so this field doesn't contribute to 
$a_{body}$.

Let us now simplify the second tern in Eq. (\ref{3}). Because one should neglect signal delay in $T_{\alpha\alpha}$ and in $T_{00}$ vise versa, we have to take it into account, let us perform expansion in the powers of $t' = -|{\bf r}-{\bf r}'|$:
\begin{eqnarray}
h_{00}({\bf r}, t) & = & -2\int\frac{T_{00}({\bf r}', t')}{|{\bf r}-{\bf r}'|}d^3{\bf r}' - 2\int\frac{\sum T_{\alpha\alpha}({\bf r}', t')}{|{\bf r}-{\bf r}'|}d^3{\bf r}' \approx
\nonumber \\
    & \approx & -2\int\frac{T_{00}({\bf r}', 0)}{|{\bf r}-{\bf r}'|}d^3{\bf r}' + 2\int\frac{\partial T_{00}({\bf r}', 0)}{\partial t}d^3{\bf r}' - \int\frac{\partial^2T_{00}({\bf r}', 0)}{\partial t^2}|{\bf r}-{\bf r}'|d^3{\bf r}' -
\nonumber \\
    & & - 2\int\frac{\sum T_{\alpha\alpha}({\bf r}', 0)}{|{\bf r}-{\bf r}'|}d^3{\bf r}'\ \ \ .
\nonumber
\end{eqnarray}
Then we have
\begin{displaymath}
\int\frac{\partial T_{00}}{\partial t}d^3{\bf r} = \int\frac{\partial T^{00}}{\partial t}d^3{\bf r} = - \int\frac{\partial T^{0\beta}}{\partial x^{\beta}}d^3{\bf r} = 0\ \ \ ,
\end{displaymath}
\begin{displaymath}
\frac{\partial^2 T_{00}}{\partial t^2} = \frac{\partial^2 T^{00}}{\partial t^2} = - \frac{\partial}{\partial t }\frac{\partial T^{0\beta}}{ \partial x^{\beta}} = - \frac{\partial}{\partial x^{\beta}}\frac{\partial^2 T^{\beta 0}}{\partial t} = 
\frac{\partial^2 T^{\beta\alpha}}{\partial x^{\alpha}\partial x^{\beta}}\ \ \ .
\end{displaymath}
Thus, the second term in Eq. (\ref{3}) is expressed as follows:
\begin{eqnarray}
\label{14}
({\bf a}_{body})_{\alpha} & = & ({\bf a}_0)_{\alpha} + \frac{1}{2}\int\frac{T_{\beta\gamma}({\bf r}, 0)}{r^3}\left( 3\delta_{\beta\gamma}x_{\alpha} + \delta_{\alpha\beta}x_{\gamma} + \delta_{\alpha\gamma}x_{\beta} - 
\frac{3x_{\alpha}x_{\beta}x_{\gamma}}{r^2} \right) d^3{\bf r} +
\nonumber \\
   & + & \int\frac{\nabla_{\alpha}T_{00}({\bf r}, 0)}{r}d^3{\bf r} =
\nonumber \\ 
   & = & ({\bf a}_0)_{\alpha} + \frac{1}{2}\int\frac{d^3{\bf r}}{r}\nabla_{\alpha}T_{\beta\gamma}({\bf r}, 0)\left( 3\delta_{\beta\gamma} - \frac{x_{\beta}x_{\gamma}}{r^2} \right) + \int\frac{\nabla_{\alpha}T_{00}({\bf r}, 0)}{r}d^3{\bf r} \equiv
\nonumber \\
   & \equiv & ({\bf a}_0)_{\alpha} + ({\bf a}_1)_{\alpha} + ({\bf a}_2)_{\alpha}
\nonumber \ \ \ ,
\end{eqnarray}
where through $({\bf a}_2)_{\alpha}$ we denoted the summand with $T_{00}$.

The tensor $T_{\alpha\beta}$ is equal to the sum of a tension tensor of the shell and field. Tensions in a shell are due to acceleration and mutual repulsion of the shell parts. As it was shown above, the former is zero. One can also neglect the 
latter because they are of the order of $q^2$ and we are only
interested in linear terms (and this means linearity in $q$). Thus, in this approximation we should take into account only $T_{\alpha\beta}$ of the electromagnetic field:
\begin{displaymath}
T_{\alpha\beta} \approx \frac{1}{4\pi}\left( \frac{1}{2}E^2 \delta_{\alpha\beta} - E_{\alpha}E_{\beta} \right)\ \ \ ,
\end{displaymath}
where ${\bf E} = {\bf E}_0 + {\bf E}_1$,
\begin{displaymath}
\left\{
\begin{array}{ccl}
{\bf E}_0 & = & E_0\hat {\bf z}  \\
{\bf E}_1 & = & \theta (r-R)\frac{q{\bf r}}{r^3} \mbox{ is the field generated by the shell} \ .
\end{array}
\right.
\end{displaymath}
Since we retain only the terms linear in $q$,
\begin{displaymath}
\begin{array}{rcl}
E^2 & \approx & E_0^2+2{\bf E}_0{\bf E}_1 \\
E_\alpha E_\beta & \approx & E_{0\alpha}E_{0\beta}+E_{0\alpha}E_{1\beta}+
E_{0\beta}E_{1\alpha}
\end{array} \ .
\end{displaymath}
The terms quadratic in $E_0$ can be made infinitesimal by removing the charge $Q$ to infinite distance (Eq. (\ref{13})).

Thus, we have the following expression for $T_{\alpha\beta}$ linear in $q$:
\begin{displaymath}
T_{\alpha\beta} = \frac{qE_0}{4\pi r^2}(\hat z_\gamma \delta_{\alpha\beta} - \hat z_\alpha \delta_{\gamma\beta} - \hat z_\beta \delta_{\alpha\gamma})\hat r_\gamma \theta (r-R)\equiv \frac{qE_0}{4\pi r^2}K_{\alpha\beta\gamma}\hat r_\gamma \theta 
(r-R)\ \ \ .
\end{displaymath}

Now ${\bf a}_1$ can be expressed in the form convenient for calculations:
\begin{displaymath}
({\bf a}_1)_\alpha = 2\pi \int^\infty_0 dr \frac{qE_0}{4\pi r^2} K_{\alpha\beta\gamma} \left\langle 3\delta_{\beta\gamma}\hat r_\delta \hat r_\gamma + \delta_{\alpha\beta} \hat r_\delta \hat r_\gamma + \delta_{\alpha\gamma} \hat r_\delta \hat 
r_\beta  - 3\hat r_\alpha \hat r_\beta \hat r_\gamma \hat r_\delta \right\rangle \theta (r-R)\ \ \ ,
\end{displaymath}
where the angular brackets $\langle\cdots\rangle $ denote averaging over angles. After this averaging, we have
\begin{eqnarray}
({\bf a}_1)_\alpha & = & \int^\infty_0 dr \frac{2aE_0}{5\pi r} \theta (r-R) (K_{\beta\beta\alpha} + \frac{1}{3}K_{\alpha\beta\beta}) =
\nonumber \\
   & = & \int^\infty_0 dr \frac{2aE_0}{5\pi r} \theta (r-R) [\hat z_{\alpha} + \frac{1}{3}(-3\hat z_{\alpha})] = 0
\nonumber \ \ \ .
\end{eqnarray}

Now let us calculate the contribution due to $T_{00}$, i.e. ${\bf a}_2$:
\begin{displaymath}
{\bf a}_2 = \nabla J, \mbox{ where}\ J= \int \frac{T_{00}({\bf r}', 0)}{|{\bf r} - {\bf r}'|}d^3 {\bf r}' = J^{shell}_{1} + J^{field}_{2}\ \ \ .
\end{displaymath}
It is obvious that the summand due to the shell, i.e. $J^{shell}_{1}$, is spherically symmetrical (the Lorentz contraction, proportional to $a^2$, is superfluous in our approximation). Therefore, it is clear that $\nabla J_1 |_{{\bf r} =0}=0$. The only 
remaining contribution is that due to the electric field, $J^{field}_2$:
\begin{displaymath}
T_{00} \approx \frac{E^2}{8\pi}, \mbox{ where } {\bf E}={\bf E}_0 + {\bf E}_1\ \ \ .
\end{displaymath}
The term linear in $q$ and ${\bf E}_0$ is equal to
\begin{displaymath}
\frac{qE_0}{4\pi r^2}(\hat{\bf r}\cdot \hat{bf z})\theta (r-R)\ \ \ .
\end{displaymath}
Expanding $1/|{\bf r} - {\bf r}'|$ in ${\bf r}$, we obtain
\begin{eqnarray}
{\bf a}_2 & = & \nabla\int\frac{qE_0}{4\pi {r'}^2}\theta(r'-R)(\hat{\bf r}'\cdot\hat{\bf z})(\frac{\hat{\bf r}'}{{r'}^2}{\bf r})d^3{\bf r}' = \nonumber \\
& = & \frac{qE_0}{4\pi}4\pi\int^{\infty}_R\frac{d{\bf r}'}{{r'}^2}\langle (\hat{\bf r}'\cdot\hat{\bf z})\hat{\bf r}'\rangle =\frac{qE_0}{3R}\hat{\bf z}
\nonumber\ \ \ .
\end{eqnarray}

Taking into account Eq. (\ref{12}), the final result is
\begin{displaymath}
{\bf a}_{body} = \frac{13M}{3R}{\bf a} = \frac{13R_g}{6R}{\bf a}\ \ \ ,
\end{displaymath}
i.e. the total effect is the entrainment of the probe in the direction of the acceleration vector.

%\begin{center}
\section{Conclusion}
%\end{center}

The main aim of this methodological paper was to elucidate the role of the energy-momentum tensor in the general relativity, taking as an example a shell moving at the constant acceleration. The paper would be useful for the development of the 
``relativistic intuition'' of the beginners in gravity. In particular,
it was demonstrated that the removal from the full EMT of the parts that seem to be unimportant from the point of view of Newton's gravity law, leads to incorrect and, 
usually, paradoxical results.

\setcounter{equation}{0}
\renewcommand{\theequation}{A.\arabic{equation}}
%\begin{center}
\section{Appendix. Tension Tensor in a Rigid Thin Shell Accelerated by a String}
%\end{center}

One can neglect tensions $T_{r\alpha}$ \cite{6}, where $\alpha = r,\ \theta ,\ \varphi $ are components of the spherical frame system with a polar axis directed along the string. Owing to the axial symmetry, only the following components of the 
tension tensor are nonzero:
\be
T_{\theta\theta} \ , \ T_{\varphi\varphi} \ \ \ .
\ee
Taking the above into account, the most general form of $T_{\alpha\beta}$ is
\be
T_{\alpha\beta} =
\left[
A(\theta)\hat \theta_{\alpha}\hat \theta_{\beta} +
B(\theta)\hat \varphi_{\alpha}\hat \varphi_{\beta}
\right]
\delta(r-R) \ ,
\ee
where $\hat \theta,\ \hat \varphi$ are unit vectors tangent to the arcs $\varphi = const$ and $\theta = const$ (see Fig. 3).

Let us consider the balance condition for forces in the accelerated frame system. The mass of shell section corresponding to the polar angles $\theta <\theta '<\pi$ (the dashed section in Fig. 3) is:
\begin{displaymath}
M(\theta)=\frac{M}{2}(1+\cos\theta) \ \ \ .
\end{displaymath}
The force acting on this section (it is obvious that in this motion only the $z$-components of the force is nonzero) is \cite{6,7}
\be
F_{\alpha} = -\int T_{\alpha\beta}(d{\bf f})_{\beta} \  \ \ ,
\ee
where $d{\bf f}$ is an element of the surface restricting given body. In our case $d{\bf f}$ is an element of the sphere on the circle $\theta = const$ within the range of the azimuth angle $(\varphi,\varphi + d\varphi )$: $\hat\theta d{\bf f}= 
-r\sin\theta drd\varphi$. Then, from Eqs. (A.2) and (A.3) we obtain
\begin{displaymath}
{\bf  F} = A(\theta)R\sin\theta\int\hat\theta d\varphi =
-2\pi R A(\theta)R\sin^2\theta\hat{\bf z}\ \ \ .
\end{displaymath}
This force is compensated by the force of inertia ${\bf F} = M(\theta)a\hat{\bf z}$. Hence 
\be
A(\theta) = -\frac{Ma}{4\pi R}\frac{(1+\cos\theta}{\sin^2\theta)} \ .
\ee
The singularity at $\theta = 0$ is related to the fact that the infinitely thin  string is attached to the sphere exactly at this point.

In order to calculate $B(\theta)$, let us consider the equilibrium of an arbitrary shell element in the radial direction (Fig. 3). The forces ${\bf F}_1$ and ${\bf F}_2$ are perpendicular to the parallels $\theta = const$, and ${\bf F}_3$ and ${\bf 
F}_4$ to the meridians $\varphi = const$. Using Eqs. (A.1) and (A.2), we can express these forces as
\begin{displaymath}
\left\{
\begin{array}{l}
{\bf F_{1,2}} = \mp\hat\theta A(\theta)R\sin\theta\Delta\varphi  \\
{\bf F_{3,4}} = \mp\hat\varphi B(\theta)R\Delta\theta
\end{array}
\right. \ \ \ .
\end{displaymath}
The radial component of the net force  is
\begin{displaymath}
\hat{\bf r}\cdot {\bf F} =
\left(
A(\theta)+B(\theta)
\right)
R\sin\theta\Delta\varphi\Delta\theta \ \ \ .
\end{displaymath}
Adding this result with the radial component of the inertia force
\begin{displaymath}
{\bf F}^{inert}_{rad}=-\frac{Ma}{4\pi }\sin\theta\cos\theta\Delta\varphi\Delta\theta
\end{displaymath}
and equating the result with zero, we obtain the equilibrium condition
\begin{displaymath}
A(\theta)+B(\theta)=\frac{Ma}{4\pi R}\cos\theta \ \ \ .
\end{displaymath}
>From this and from Eq. (A.4), we obtain
\begin{displaymath}
B(\theta)=\frac{Ma}{4\pi R}\frac{(1+\cos\theta +\cos\theta\sin^2\theta)}
{\sin^2\theta} \ \ \ .
\end{displaymath}
>From this result and Eqs. (A.2) and (A.4), we derive Eq. (\ref{8}).

\end{document}